# Estimating and decomposing most productive scale size in parallel DEA networks with shared inputs: A case of China's Five-Year Plans

Saeed Assani[1,2*] · Jianlin Jiang[1] · Ahmad Assani[3] · Feng Yang[2]

**Abstract.** Attaining the optimal scale size of production systems is an issue frequently found in the priority questions on management agendas of various types of organizations. Determining the most productive scale size (MPSS) allows the decision makers not only to know the best scale size that their systems can achieve but also to tell the decision makers how to move the inefficient systems onto the MPSS region. This paper investigates the MPSS concept for production systems consisting of multiple subsystems connected in parallel. First, we propose a relational model where the MPSS of the whole system and the internal subsystems are measured in a single DEA implementation. Then, it is proved that the MPSS of the system can be decomposed as the weighted sum of the MPSS of the individual subsystems. The main result is that the system is overall MPSS if and only if it is MPSS in each subsystem. MPSS decomposition allows the decision makers to target the non-MPSS subsystems so that the necessary improvements can be readily suggested. An application of China's Five-Year Plans (FYPs) with shared inputs is used to show the applicability of the proposed model for estimating and decomposing MPSS in parallel network DEA. Industry and Agriculture sectors are selected as two parallel subsystems in the FYPs. Interesting findings have been noticed. Using the same amount of resources, the Industry sector had a better economic scale than the Agriculture sector. Furthermore, the last two FYPs, 11[th] and 12[th], were the perfect

This work is supported by the National Natural Science Foundation of China (No. 71631006, 11571169).

* Saeed Assani
  saeedassani@nuaa.edu.cn
  Tel: +8615077820900
  Jianlin Jiang
  jiangjianlin@nuaa.edu.cn
  Ahmad Assani
  ahmad.assani@hs-karlsruhe.de
  Feng Yang
  fengyang@ustc.edu.cn
[1] College of Science, Nanjing University of Aeronautics and Astronautics, Nanjing 210016, China
[2] School of Management, University of Science and Technology of China, Hefei 230026, China
[3] Faculty of Computer Science and Business Computer Systems, Karlsruhe University of Applied Science, Karlsruhe, 76133, Germany

two FYPs among the others.

**Keywords:** Data envelopment analysis • Most productive scale size • Parallel Network • Industry • Agriculture • Five-Year Plans

# 1 Introduction

Data envelopment analysis (DEA) is a mathematical method for measuring the relative efficiency of decision making units (DMUs) which may have multiple inputs and outputs (Charnes, Cooper, & Rhodes, 1978). DEA was accorded this name because of the way it envelops the DMUs to identify an efficiency frontier that is used to evaluate the DMUs. On the efficient frontier, there is a unit at which the average productivity of the DMU inputs and outputs mix is maximized. This point is called the most productive scale size (MPSS), and it is first introduced to standard DEA by (Rajiv D. Banker, 1984).

(Joe Zhu & Zhao-Han Shen, 1995) showed that the MPSS concept can always be used to estimate RTS without any adjustments unless a set of efficient DMUs exhibit linear dependency, i.e., it is the DMU itself that causes the MPSS concept not to work. Also, they pointed out that the MPSS concept itself is independent of assuming a linear production function in the CCR model. Cooper et al. (1996) proposed a measure of scale which is "dimensionless" (i.e., it does not depend on the units of measure used). (Zhu, 2000) gave a further discussion on linear production functions and DEA, where MPSS was the main research discussion. Later, (Rajiv D. Banker, Cooper, Seiford, Thrall, & Zhu, 2004) discussed RTS in DEA for each of the presently available types of models. In recent years, (Wang & Lan, 2013) defined the MPSS concept from a pessimistic perspective. Then they used a double frontier approach to integrate the optimistic and pessimistic measures of MPSS in one term. (Lee, 2016) proposed a multi-objective mathematical program with DEA constraints to set an efficient target that shows a trade-off between the MPSS benchmark and a potential demand fulfillment benchmark. The classic data envelopment analysis requires that the values for all inputs and outputs be known exactly. However, this assumption may not be true, because data in many real applications cannot be precisely measured. One of the important methods to deal with imprecise data is considering stochastic data in DEA. Therefore, Khodabakhshi (2009) studied the most productive scale size by considering stochastic data in DEA. To that end, he extended the work of (Jahanshahloo & Khodabakhshi, 2003) in stochastic data envelopment analysis. To solve the stochastic model, a deterministic equivalent is obtained. Although the deterministic equivalent is non-linear, it can be converted to a quadratic program. (Eslami, Khodabakhshi, Jahanshahloo, Hosseinzadeh Lotfi, & Khoveyni, 2012) dealt with a realistic decision problem that contains fuzzy constraints and uncertain information (stochastic data) that most productive scale size (MPSS) is estimated in imprecise-chance constrained DEA model. (Davoodi, Zarepisheh, & Rezai, 2014) introduced a notion of the nearest MPSS pattern, which yields the closest MPSS pattern compared to all others. By the aid of this pattern, a unit would be able to reach its optimal size more easily and by small changes in its inputs and outputs. (Sahoo, Khoveyni, Eslami, & Chaudhury, 2016) proposed a non-radial DEA model to determine the MPSS and RTS of a DMU in the presence of negative data. (Dwi Sari, Angria S, Efendi, & Zarlis, 2018) introduced a new MPSS model can deal with integer value data.

Despite the importance of MPSS and abundance of publications on the efficiency of network DEA, MPSS has not been enough addressed in the network DEA literature. Recently (Assani, Jiang, Cao, & Yang, 2018) introduced MPSS to multi-stage systems which are connected in series and proposed new models to measure the MPSSs of the system and the internal stages. Also, they developed an approach to derive the MPSS projections of non-MPSS DMUs.

Another type of network DEA is a parallel network where all processes are operated independently. Several studies used this network to measure the overall efficiency of the evaluated system and its subsystems (Amirteimoori & Yang, 2014; An, Yang, Chu, Wu, & Zhu, 2017; Du, Zhu, Cook, & Huo, 2015; Gong, Zhu, Chen, & Cook, 2018; Hosseini & Stefaniec, 2019; C. Kao, 2012; Lei, Li, Xie, & Liang, 2014; F. Yang, Du, Liang, & Yang, 2014). For example, a government Five-Year Plan (FYP) focuses on several sectors during the same period, and each sector can be considered as a subsystem of a parallel FYP. It would be interesting if the policymakers could measure the productivity scale size of the FYPs and their sectors, subsystems.

As it is known, China is one of the first countries, which used the FYP system in their national planning. Since 1953, 13 series of social and economic development initiatives have been issued mapping strategies for economic development, setting growth targets, and launching reforms. Each FYP has its own highlighted sectors, additionally to the main sectors. Most FYPs focused on service sectors such as health-care, education, and transportation and production sectors such as the economy, industry, and agriculture. The policymakers consider those sectors as a parallel structure and work on investing the resources to achieve higher revenue, either social or financial, from each sector. Rapid development in Chinese industrial sectors pushes the Chinese government to work on creating new policy indicators for future Five-Year Plans. To accomplish that, we evaluate and measure the productivity scale size of the selected industry sectors along previous FYPs. This evaluation can help the decision makers to identify the sectors that need much interests and investment to achieve the best scale size in future FYPs. It would be interesting if the policymakers could measure the productivity scale size of the FYPs and their sectors, subsystems.

In DEA literature, several papers evaluated the performance and productivity of government planning strategies. (Bi, Wang, Yang, & Liang, 2014) presented a non-radial DEA model with multidirectional efficiency analysis (MEA) involving undesirable outputs to measure the regional energy and environmental efficiency of the transportation sector during the 11$^{th}$ China's FYP, period 2006-2010. (Wu, Shi, Xia, & Zhu, 2014) used the super-efficiency DEA window analysis to evaluate the circular efficiency of Chinese regions during the 11$^{th}$ China's FYP. (M. Yang & Yang, 2016) evaluated and compared the environmental-adjusted energy productivity of 15 energy-intensive industries during the 10$^{th}$ and 11$^{th}$ FYPs. (Li, Jiang, Mu, & Yu, 2018) measured the efficiency and the possible saving energy of the agricultural sector of Chinese provinces from 1997 to 2014.

However, either most of these studies used traditional DEA models without considering the relationships may exist between the internal processes, mentioned one sector for evaluation, or they are done at the provincial level. Furthermore, a little work in the literature that focused on the productivity scale.

This study makes contributions to the methodological and practical sides. At the methodological side, we propose a relational model can measure both the MPSS of a parallel network and the subsystems' MPSSs as well. At the practical side, this study is the first that measures the most productive scale size of government plans using parallel DEA network considering multiple sectors for evaluation in one DEA implementation.

This paper is organized as follows. In the next section, a relational model for measuring the MPSS of the parallel network is proposed with an illustrative data example. A real-life application of China's Five-Year Plans is introduced in Section 3. Conclusions are reported in Section 4.

## 2  MPSS for parallel production systems

Assani et al. (2018) generalized the MPSS concept from black-box DEA into multi-stage DEA. In their models, they have intermediate measures connecting the internal stages. These intermediate measures have been adjusted non-radially in the proposed MPSS models while the inputs and outputs are radially adjusted. In our study, we do not consider any intermediate measures between the subsystems. In the following subsection, we propose a relational model to measure the MPSS of the evaluated system and its subsystems.

### 2.1  Relational MPSS model

Consider a standard parallel system in which $h$ subsystems, processes, are connected in parallel to form a network system (see Figure 4-1). Let $X_{ij}^{(t)}$ ($i = 1,2,\ldots,m$) be the $i$th input and $Y_{rj}^{(t)}$ ($r = 1,2,\ldots,s$) the $r$th output of subsystem $t$ ($t = 1,2,\ldots,h$) for DMU j ($j = 1,2,\ldots,n$). The sum of the $i$th input for all subsystems is equal to the $i$th input of the system of DMU j, i.e. $\sum_{t=1}^{h} X_{ij}^{(t)} = X_{ij}$. This also applies to outputs; that is, $\sum_{t=1}^{h} Y_{rj}^{(t)} = Y_{rj}$.

In our proposed model, the MPSS of the whole system and the internal processes are measured in a single DEA implementation.

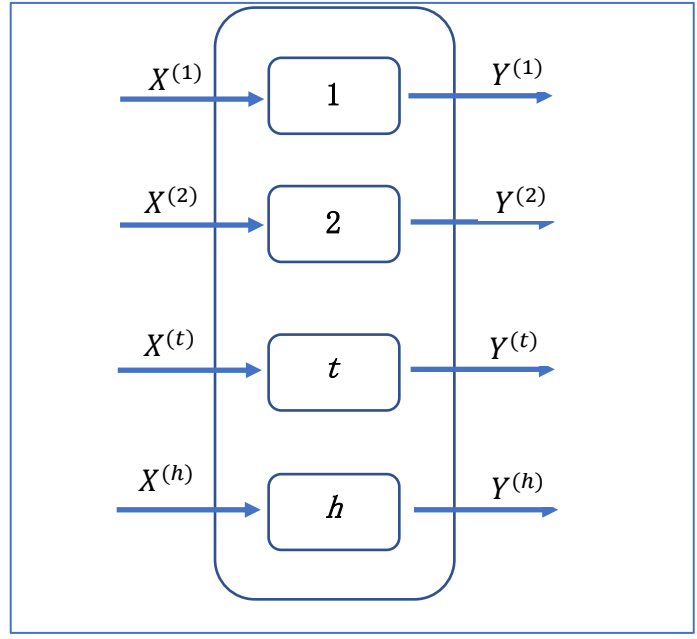

**Figure 1 Classical parallel structure**

The proposed relational MPSS model for a classical parallel network can be expressed as follows:

$$MPSS^{S*} = \max \emptyset - \theta \tag{1}$$

$$s.t. \begin{cases} \sum_{j=1}^{n} \lambda_j^t X_{ij}^t \leq \theta^t X_{io}^t, i = 1,2,\dots,m, t = 1,2,\dots,h \\ \sum_{j=1}^{n} \lambda_j^t Y_{rj}^t \geq \emptyset^t Y_{ro}^t, r = 1,2,\dots,s, t = 1,2,\dots,h \\ \sum_{j=1}^{n} \lambda_j^t = 1, t = 1,2,\dots,h \\ \lambda_j^t, \theta^t, \emptyset^t \geq 0 \; \forall j, t = 1,2,\dots,h \end{cases} \tag{1.1}$$

$$\begin{cases} \sum_{j=1}^{n} \mu_j X_{ij} \leq \theta X_{io}, i = 1,2,\dots,m \\ \sum_{j=1}^{n} \mu_j Y_{rj} \geq \emptyset Y_{ro}, r = 1,2,\dots,s \\ \sum_{j=1}^{n} \mu_j = 1 \\ \mu_j, \theta, \emptyset \geq 0, \forall j \end{cases} \tag{1.2}$$

$$\begin{cases} \theta = \sum_{t=1}^{h} \omega^t \theta^t \\ \emptyset = \sum_{t=1}^{h} \omega^t \emptyset^t \end{cases} \qquad (1.3)$$

In constraints (1.1), each process $t$ has its own set of intensity coefficients, $\lambda_j^t, j = 1,2,\ldots,n$ and distance measures $\theta^t, \emptyset^t$ as well. $\theta^t, \emptyset^t$ are scalars representing expansion or contraction factors applied to the inputs and outputs of process $t$ in the evaluated DMU$_o$.

The constraints (1.2) represent the system constraints, where the inputs and outputs of processes are aggregated to the total inputs and outputs of the system. Based on the efficiency decomposition concept of parallel networks (see (Chiang Kao, 2014)), We use $\theta, \emptyset$ as a weighted average of the corresponding distance measures of the internal processes as in constraints (1.3).

The variables $\omega^t, t = 1,2,\ldots,h$ are weights that reflect the preference over the $h$ processes' performances and are selected by the decision maker. However, these three variables are exogenous variables that cannot be determined by the proposed model. In this study, we set $\omega^t = 1, \forall t$ as we assume that all processes are equal in importance to the decision maker.

As we mentioned above, our proposed MPSS model can measure the overall and internal MPSS scores of a parallel network. The objective function value of model (1) is the overall MPSS of the parallel network while the amount $\emptyset^{t*} - \theta^{t*}$ is the MPSS score of process $t$.

**Definition 1** DMU$_o$ is (overall) MPSS if and only if the optimal objective function value of model (1) is zero.

Similar to Definition 1, we define the MPSS of process $t$ for DMU$_o$ as follows.

**Definition 2** DMU$_o$ is MPSS in process $t$ if and only if $\boldsymbol{\emptyset^{t*} - \theta^{t*} = 0}$ in the optimal solution of model (1).

## 2.2 The relationship between the system MPSS and the subsystems MPSSs

In this subsection, the relationship between the MPSS of the system and those of the individual subsystems is identified. It can be seen that the system MPSS can be decomposed into individual stages. With the aid of MPSS decomposition, the decision makers can identify the source of non-MPSS state and find out where the adjustments should be made to improve the performance of the evaluated DMU. The following theorems address this decomposition.

**Theorem 1** MPSS of a parallel network system is the weighted sum of the MPSS of the internal processes; that is, $MPSS^{s*} = \sum_{t=1}^{h} \omega^t MPSS^{t*}$.

**Proof**:

Based on the objective function value of model (1) and constraints (1.3), $MPSS^{s*} = \emptyset^* - \theta^*$ and $\theta = \sum_{t=1}^{h} \omega^t \theta^t$, $\emptyset = \sum_{t=1}^{h} \omega^t \emptyset^t$, respectively. As a result, $MPSS^{s*} = \emptyset^* - \theta^* = \sum_{t=1}^{h} \omega^t \emptyset^{t*} - \sum_{t=1}^{h} \omega^t \theta^{t*} = \sum_{t=1}^{h} \omega^t (\emptyset^{t*} - \theta^{t*}) = \sum_{t=1}^{h} \omega^t MPSS^{t*}$.

Based on Theorem 1, we have the following theorem.

**Theorem 2** A DMU is overall MPSS if and only if it is MPSS in each process.

**Proof**. When a DMU is overall MPSS, its MPSS score is zero. According to Theorem 1 and the non-negativity of $MPSS^{t*}$ ($t = 1,2,...,h$), each process has a score equal to zero, and this means that all processes are MPSS. Conversely, when each process is MPSS, based on Theorem 1, the system MPSS score is zero, and thereby the DMU is overall MPSS.

The proposed model and theoretic analysis in this section provided some insights into the MPSS in parallel network DEA, including measuring the MPSS of the system and sub-systems and establishing the relationship between them. In the following, we give an illustrative data example to show the applicability of the above discussion.

## 2.3 Illustrative data example

The representative dataset contains five DMUs, as shown in Table 1. Each DMU has a parallel structure (*h*=2). Each subsystem consumes one input and produces one output.

**Table 1 Illustrative dataset for MPSS measurement**

| DMU | Subsystem I | | Subsystem II | | $MPSS^I$ | $MPSS^{II}$ | $MPSS^S$ |
|---|---|---|---|---|---|---|---|
| | X1 | Y1 | X2 | Y2 | | | |
| A | 2 | 2 | 2 | 3 | 1 | 1.25 | 2.25 |
| B | 3 | 5 | 1 | 4 | 0 | 0 | 0 |
| C | 5 | 2 | 1.5 | 6 | 1.90 | 0 | 1.90 |
| D | 4 | 4 | 2 | 3 | 0.50 | 1.25 | 1.75 |
| E | 2 | 1 | 4 | 2 | 3.50 | 2.62 | 6.12 |

Applying model (1), the MPSS of the system and the subsystems can be obtained, as shown in Table 1. According to Theorem 1, the MPSS of the system for each DMU is the sum of the MPSS values of the subsystems (here we assume the weights $\omega^1 = \omega^2 = 1$). Using DMU *B* to explain this, the MPSS of the system is zero, which is precisely the sum of the MPSS values of subsystem I (0), subsystem II (0). Similarly, for DMUs *A*, *C*, *D*, and *E* we have 2.25=1+1.25, 1.90=1.90+0, 1.75=0.50+1.25, and 6.12=3.50+2.62, which satisfy this theorem. With the MPSS decomposition, we can determine the MPSS of each subsystem and find its contribution to the system's MPSS.

Theorem 2 reveals that $DMU_o$ is overall MPSS if and only if it is MPSS in each subsystem, which can also be observed from Table 1. Only DMU *B* is overall MPSS because the MPSS scores of the system and the two subsystems are zero, while the other DMUs are not MPSS. Note that DMU *C* is MPSS in the second subsystem, while it is not MPSS in the first subsystem. Based on this decomposition, the decision makers can make the right decision to improve the performance of the first subsystem for DMU *C* to achieve the most productive scale size.

## 3   China's Five-Year Plans

Since 1953, 13 series of social and economic development initiatives have been issued mapping strategies for economic development, setting growth targets, and launching reforms. Each FYP has its own highlighted sectors, additionally to the main sectors. Rapid development in Chinese industrial sectors pushes the Chinese government to work on creating new policy indicators for future Five-Year Plans. To accomplish that, we evaluate and measure the productivity of the selected industry sectors along previous FYPs. This evaluation can help the decision makers to identify the sectors that need much interests and investment to achieve the best scale size in future FYPs.

In DEA literature, there is a lake in the studies that concerned in the productivity scale size of governmental plans and strategies in the main sectors such as industry, transport, environment, agriculture, etc. Most studies are at the efficiency level. In this study, we consider two sectors as two subsystems in the evaluated FYP. The first sector is the Industry, and the second is Agriculture.

Our motivation to provide the Five-Year Plans application is two-fold. On the one hand, we aimed to apply our new MPSS models of parallel networks to measure the productivity scale size of the governmental plans based on specific sectors, Industry and Agriculture. These two sectors were from the top critical indicators in the last ten FYPs. We chose these two sectors based on the history and the present of the Chinese economy. As we know that China was an agricultural country, and now China plays a global role in the industry sector. The data of the two sectors also are available and collectible. On the other hand, we aimed to tell the scholars that DEA can play a more significant role for short-term and long-term planning by evaluating the government work in terms of some sectors and policy indicators.

We believe that the two sectors, Industry and Agriculture, can be treated as a parallel system because any government aims to develop all the main sectors together with different rate of importance for each sector. We evaluate the industry and the agriculture sectors based on the given inputs and the produced outputs. Another combination of inputs and outputs will lead to different results and assessments. At the current inputs and outputs, the interaction may occur only in the shared inputs, and that is common in DEA application, and thus we made the inputs shared between the two sectors. We assumed that we would not consider any intermediate measures between the subsystems; otherwise it will not be called a parallel system.

Assessing the most productive scale size of these plans based on these sectors will help the policymakers to know the place of inefficiency and the amendment needed to improve these sectors. MPSS decomposition will be proposed to help the decision makers to find the sector that does not achieve the most productive scale size in each year of the FYP.

## 3.1 Specification of input and output variables

As we mentioned above, there are 13 FYPs since China started this program in its government planning. Due to the difficulty in getting some data in the so-far FYPs, we only focus on the last ten FYPs. We collected the data since the year 1966 until the end of 2015, that is, 50 years represent ten FYPs. The full data are displayed in Appendix A. We consider each year as a DMU and thus all years will be evaluated based on the same frontier. By linking every five years with the corresponding FYP, we get some interesting findings.

Inputs and outputs data of chosen sectors were collected from two sources (National Bureau of Statistics of China, World Bank National). Due to the difficulty in collecting some data from the so far years of first and second FYPs of China, we only consider the last ten FYPs. Table 2 provides the descriptive statistics of inputs/outputs for the ten FYPs.

**Inputs:** all subsystems share the same inputs (Population, GDP per capita, government final consumption expenditure) which are considered the generators of the national economies. GDP and population are two critical indicators of economic. The population can generate and raise the GDP level by their productive works (Gutierrez, Glassman, Steven, & Marcuss, 2009; Landefeld, Seskin, & Fraumeni, 2008). Government final consumption expenditure is one of the primary indicators which support economic growth. As it is proved, there is a dynamic causal relationship between government expenditure and economic growth (Landau, 1983; Odhiambo, 2015).

GDP per capita as it is known is the gross domestic product divided by midyear population. Data are in current U.S. dollars. General government final consumption expenditure or so-called general government consumption (GFC) includes all current government expenditures for purchases of goods and services.

**Outputs:** each subsystem produces its value-added. Namely, Industry value-added and Agriculture value-added.

The value-added of each sector is the net output of this sector after adding up all outputs and subtracting intermediate inputs (Roblek, Meško, & Krapež, 2016; World Bank, 2000, 2012).

**Table 1 Summary of inputs and outputs descriptive statistics of the last ten FYPs**

|  | Shared Inputs | | | Outputs | |
| --- | --- | --- | --- | --- | --- |
|  | **Population** | **GDP per capita** | **GFC** | **Industry VA** | **Agriculture VA** |
| **Min** | 735400000 | 91.47271831 | 7819481680 | 22040783167 | 28523844342 |
| **Max** | 1371220000 | 8069.213024 | 1.54615E+12 | 4.52895E+12 | 9.77311E+11 |
| **Mean** | 1112552600 | 1385.733979 | 2.46836E+11 | 8.10376E+11 | 2.09068E+11 |
| **S.D.** | 191765333 | 2139.63222 | 3.94325E+11 | 1.28523E+12 | 2.55812E+11 |

## 3.2 Specification of China's FYPs MPSS model

Since the inputs of the two sectors, Industry and Agriculture, are shared, model (1) should be revised.

Let a parameter $\alpha$ $(0 < \alpha < 1)$ denote the proportion of inputs to be dedicated to the Industry subsystem. Then, the overall inputs $(X_j)$ are divided into two parts $X_j^I$ and $X_j^A$ for the two subsystems, Industry and Agriculture, respectively, as follows:

$$X_j^I = \alpha X_j \text{ and } X_j^A = (1-\alpha)X_j, \ \forall j = 1,2,\ldots,n, \tag{2}$$

where $\alpha$ is a parameter. If $\alpha = 0$, it means that all inputs are consumed by the Agriculture sector. On the contrary, $\alpha = 1$ means that all inputs $X_j$ are consumed by the Industry sector. Since Industry and Agriculture are primary sectors of all China's FYPs, inputs on each side cannot be zero. That is, each sector consumed some of the overall inputs to obtain the nonzero outputs on each side. Thus, the parameter of $\alpha$ is in an interval range of $\alpha \in (0,1)$ (See Figure 2).

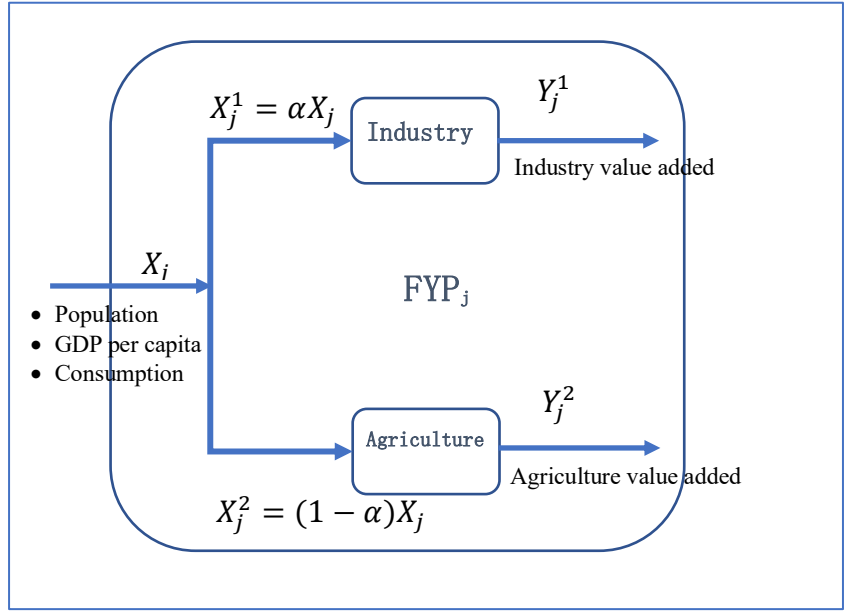

**Figure 1 A parallel system of FYP with shared inputs**

Substituting (2) in (1), we obtain the MPSS model for China's FYPs as follows:

$$Overall\ MPSS = \text{Max}\ \emptyset - \theta \tag{3}$$

$Subsystem\ 1: Industry$

$$s.t. \sum_{j=1}^{n} \lambda_j^I \alpha X_{ij} \leq \theta^I \alpha X_{io}, i = 1,2,\ldots,3$$

$$\sum_{j=1}^{n} \lambda_j^I Y_j^I \geq \emptyset^I Y_o^I$$

$$\sum_{j=1}^{n} \lambda_j^I = 1$$

$$0 < \alpha < 1, \lambda_j^I, \theta^I \geq 0, j = 1,2,\ldots,n$$

$Subsystem\ 2: Agriculture$

$$\sum_{j=1}^{n} \lambda_j^A (1-\alpha) X_{ij} \leq \theta^A (1-\alpha) X_{io}, i = 1,2,\ldots,3$$

$$\sum_{j=1}^{n} \lambda_j^A Y_j^A \geq \emptyset^A Y_o^A$$

$$\sum_{j=1}^{n} \lambda_j^A = 1$$

$$0 < \alpha < 1, \lambda_j^A, \theta^A \geq 0, j = 1,2,\ldots,n$$

Syetem's constraints

$$\sum_{j=1}^{n} \mu_j X_{ij} \leq \theta X_{io}, i = 1,2,\ldots,3$$

$$\sum_{j=1}^{n} \mu_j Y_j \geq \emptyset Y_o$$

$$\sum_{j=1}^{n} \mu_j = 1$$

$$\mu_j, \theta, \emptyset \geq 0, \forall j$$

$$\theta = \omega^I \theta^I + \omega^A \theta^A$$

$$\emptyset = \omega^I \emptyset^I + \omega^A \emptyset^A.$$

Model (3) is a nonlinear program because of variable $\alpha$. To solve this model, we use a heuristic search approach to transform it into a linear program. Here we vary the value $\alpha$ in the range $(0,1)$. To do that we choose epsilon=0.1, and $\alpha = k * epsilon$.

$$kmax = \left(\frac{1}{epsilon}\right) - 1 = 9$$

Based on this discussion, we run model (3) for the values $k = 1,2,\ldots,9$.

Epsilon can be selected as a small positive value, such as 0.1, 0.01, or 0.001. If we chose epsilon as 0.01, we would have 99 intervals for the shared inputs. That is unlogic because for the first interval the industry will get 1% of the shared inputs and the agriculture will get 99% of the inputs. In real life, the inputs (population, GDP, and government consumption) for two main sectors such as industry and agriculture will not vary significantly in such (99%:1%). Therefore, we decided to select epsilon value as 0.1 and thus we will have nine intervals of the shared inputs. The maximum difference in sharing the inputs will have (90%: 10%) or (10%:90%).

## 3.3 Results and discussion

Applying model (3) to the data set of China's FYPs, the overall, Industry, and Agriculture MPSS scores are reported in Tables (3). For each year, nine partitions of the shared inputs are chosen.

**Table 2 Overall MPSS for k=9 partitions of shared inputs**

| FYPs | Years | Overall MPSS scores, epsilon=0.1, nine partitions to the inputs | | | | | | | | |
|---|---|---|---|---|---|---|---|---|---|---|
| | | k=1 | k=2 | k=3 | k=4 | k=5 | k=6 | k=7 | k=8 | k=9 |
| 3 | 1966 | 0.2759 | 0.0820 | 0.0820 | 0.0820 | 0.0820 | 0.0820 | 0.0820 | 0.0820 | 0.0980 |
| | 1967 | 15.485 | 15.750 | 15.480 | 15.480 | 15.480 | 15.480 | 15.480 | 15.480 | 15.480 |
| | 1968 | 28.418 | 31.781 | 28.418 | 28.418 | 31.781 | 28.418 | 20.138 | 31.781 | 20.138 |
| | 1969 | 17.258 | 18.498 | 17.199 | 18.439 | 17.199 | 17.199 | 17.199 | 17.199 | 5.5956 |
| | 1970 | 0.1673 | 0.1618 | 0.1618 | 0.1618 | 0.1618 | 0.1618 | 0.1618 | 0.1618 | 0.1666 |
| 4 | 1971 | 0.1421 | 0.2507 | 0.2507 | 2.6852 | 2.6852 | 2.6852 | 2.6852 | 2.6852 | 0.2507 |
| | 1972 | 0.2194 | 0.2141 | 0.2141 | 0.2141 | 0.2141 | 1.6879 | 1.6879 | 1.6879 | 1.6879 |
| | 1973 | 0.0484 | 0.0484 | 0.0458 | 0.0458 | 0.0458 | 0.0458 | 0.0458 | 0.0458 | 0.0458 |
| | 1974 | 0.1155 | 0.1166 | 0.1133 | 1.5106 | 1.5106 | 1.5106 | 1.5106 | 1.5106 | 1.5106 |
| | 1975 | 0.0000 | 0.0000 | 0.0000 | 0.0000 | 0.0000 | 0.0000 | 0.0000 | 0.0000 | 0.0000 |
| 5 | 1976 | 0.1079 | 0.1079 | 0.0934 | 0.9933 | 0.9933 | 0.9933 | 0.9933 | 0.9933 | 0.9933 |
| | 1977 | 0.1413 | 0.1315 | 0.0943 | 0.0943 | 0.0943 | 0.0943 | 0.0943 | 0.0943 | 0.0943 |
| | 1978 | 0.1100 | 0.2089 | 0.2049 | 0.2049 | 0.2049 | 0.2049 | 0.2049 | 0.2049 | 0.2049 |
| | 1979 | 2.0089 | 0.1775 | 2.0089 | 0.0854 | 0.0824 | 0.0824 | 0.0824 | 0.0824 | 0.0824 |
| | 1980 | 1.1031 | 0.1459 | 0.1798 | 0.1798 | 0.1798 | 0.1761 | 0.1761 | 0.1761 | 0.1761 |
| 6 | 1981 | 1.1652 | 0.1260 | 0.1379 | 0.1528 | 0.1528 | 0.1528 | 0.1528 | 0.1528 | 0.1528 |
| | 1982 | 2.0115 | 0.1099 | 2.0115 | 2.0115 | 0.0230 | 0.0230 | 0.0230 | 0.0230 | 0.0230 |
| | 1983 | 0.8806 | 1.4087 | 0.8806 | 0.0481 | 0.0481 | 0.0481 | 0.0481 | 0.0481 | 0.0481 |
| | 1984 | 5.3650 | 2.6671 | 1.9106 | 1.9106 | 1.9106 | 1.9106 | 1.9106 | 1.9106 | 1.9106 |
| | 1985 | 1.1758 | 0.6042 | 0.0343 | 0.0343 | 0.0343 | 0.0343 | 0.0343 | 0.0343 | 0.0343 |
| 7 | 1986 | 0.8296 | 0.8279 | 0.8279 | 0.1208 | 0.1208 | 0.1208 | 0.1208 | 0.1208 | 0.1208 |
| | 1987 | 0.0799 | 0.0799 | 0.0799 | 0.0799 | 0.0799 | 0.0477 | 0.0477 | 0.0477 | 0.0799 |
| | 1988 | 0.7721 | 0.7721 | 0.7721 | 0.7721 | 0.7721 | 0.7565 | 0.7565 | 0.7565 | 0.7721 |
| | 1989 | 0.5185 | 0.0961 | 0.0921 | 0.0921 | 0.0921 | 0.0921 | 0.0921 | 0.0921 | 0.0921 |
| | 1990 | 1.0197 | 1.0197 | 1.0197 | 1.0197 | 1.0197 | 1.0197 | 1.0197 | 1.0197 | 0.1119 |
| 8 | 1991 | 1.1324 | 1.1324 | 1.1232 | 1.1232 | 1.1232 | 1.1232 | 1.1232 | 1.1232 | 1.1232 |
| | 1992 | 0.8791 | 0.8791 | 0.8729 | 0.8729 | 0.8729 | 0.8729 | 1.0035 | 1.0035 | 1.0035 |
| | 1993 | 0.4207 | 0.4207 | 0.4207 | 0.4207 | 0.1522 | 0.1522 | 0.2969 | 0.2969 | 0.2969 |
| | 1994 | 0.2256 | 0.0784 | 0.0721 | 0.0721 | 0.0721 | 0.0721 | 0.0661 | 0.1093 | 0.1093 |
| | 1995 | 0.0243 | 0.0243 | 0.0143 | 0.0143 | 0.0143 | 0.0000 | 0.0000 | 0.0000 | 0.0000 |
| 9 | 1996 | 0.0597 | 0.0597 | 0.0597 | 0.0190 | 0.0000 | 0.0000 | 0.0000 | 0.0000 | 0.0000 |
| | 1997 | 0.3144 | 0.3144 | 0.3144 | 0.0216 | 0.0216 | 0.0216 | 0.0216 | 0.0216 | 0.0216 |

| | | | | | | | | | |
|---|---|---|---|---|---|---|---|---|---|
| | **1998** | 0.0391 | 0.0391 | 0.0295 | 0.0295 | 0.0295 | 0.0295 | 0.0295 | 0.0295 | 0.0295 |
| | **1999** | 0.0804 | 0.0804 | 0.0658 | 0.0658 | 0.0658 | 0.0658 | 0.0047 | 0.0047 | 0.0047 |
| | **2000** | 0.1236 | 0.1275 | 0.1094 | 0.1094 | 0.1094 | 0.1094 | 0.0492 | 0.0492 | 0.0492 |
| | **2001** | 0.1398 | 0.1111 | 0.1111 | 0.1111 | 0.1111 | 0.1111 | 0.1111 | 0.0547 | 0.0547 |
| | **2002** | 0.1565 | 0.1199 | 0.1199 | 0.1199 | 0.1199 | 0.1199 | 0.1199 | 0.0664 | 0.0664 |
| **10** | **2003** | 0.1684 | 0.1481 | 0.1481 | 0.1481 | 0.1481 | 0.1481 | 0.1481 | 0.0947 | 0.0947 |
| | **2004** | 0.0220 | 0.0000 | 0.0356 | 0.0356 | 0.0356 | 0.0356 | 0.0356 | 0.0356 | 0.0322 |
| | **2005** | 0.0000 | 0.0000 | 0.0852 | 0.0852 | 0.0852 | 0.0852 | 0.0852 | 0.0852 | 0.0808 |
| | **2006** | 0.0000 | 0.0000 | 0.1141 | 0.1141 | 0.1141 | 0.1141 | 0.1141 | 0.1141 | 0.1141 |
| | **2007** | 0.0000 | 0.0000 | 0.0000 | 0.0000 | 0.0000 | 0.0752 | 0.0752 | 0.0752 | 0.0752 |
| **11** | **2008** | 0.0000 | 0.0000 | 0.0000 | 0.0000 | 0.0000 | 0.0000 | 0.0000 | 0.0009 | 0.0009 |
| | **2009** | 0.0000 | 0.0000 | 0.0000 | 0.0000 | 0.0000 | 0.0000 | 0.0000 | 0.0000 | 0.0253 |
| | **2010** | 0.0000 | 0.0000 | 0.0000 | 0.0000 | 0.0000 | 0.0000 | 0.0000 | 0.0000 | 0.0021 |
| | **2011** | 0.0000 | 0.0000 | 0.0000 | 0.0000 | 0.0000 | 0.0000 | 0.0000 | 0.0000 | 0.0000 |
| | **2012** | 0.0000 | 0.0000 | 0.0000 | 0.0000 | 0.0000 | 0.0000 | 0.0000 | 0.0000 | 0.0000 |
| **12** | **2013** | 0.0000 | 0.0000 | 0.0000 | 0.0000 | 0.0000 | 0.0000 | 0.0000 | 0.0000 | 0.0000 |
| | **2014** | 0.0000 | 0.0000 | 0.0000 | 0.0000 | 0.0000 | 0.0000 | 0.0000 | 0.0000 | 0.0000 |
| | **2015** | 0.0000 | 0.0000 | 0.0000 | 0.0000 | 0.0000 | 0.0000 | 0.0000 | 0.0000 | 0.0000 |
| **No.** | | 12 | 13 | 10 | 10 | 11 | 11 | 11 | 10 | 8 |
| **Mean** | | 1.6641 | 1.5784 | 1.5202 | 1.5640 | 1.5609 | 1.5233 | 1.3606 | 1.5911 | 1.0611 |
| **Min** | | 0.0000 | 0.0000 | 0.0000 | 0.0000 | 0.0000 | 0.0000 | 0.0000 | 0.0000 | 0.0000 |
| **Max** | | 28.418 | 31.780 | 28.418 | 28.418 | 31.780 | 28.418 | 20.138 | 31.780 | 20.138 |

Table 3 reports the MPSS score for 50 years of data of the last ten FYPs. The first column displays the FYP's number. The next columns show the MPSS scores for nine partitions of shared inputs. These partitions can give us some insights on how should the resources be reallocated to achieve the most productive scale size. For example, the 3rd, 5th, 6th, and 7th FYPs have not been MPSS in any years even with varying the ratio of shared inputs between the two sectors, Industry and Agriculture. In contrast, the 4th, 8th, 9th, 10th, 11th, and 12th FYPs have at least one MPSS year.

The 4th FYP is MPSS only in one year, 1975 whatever the ratio of the shared inputs is. The next MPSS year is 1995, in the 8th FYP. Only it is MPSS when the inputs are shared in these ratios (60 Industry: 40 Agriculture; 70 Industry: 30 Agriculture; 80 Industry: 20 Agriculture; 90 Industry: 10 Agriculture). In a similar situation, the 9th FYP in the year 1996 is also MPSS when the Industry sector shares at least 50% of the inputs. In the 10th FYP, only two years, 2004 and 2005, are MPSS when the Agriculture sector shares at least 80% of the inputs.

Very notable improvements are noticed at the beginning of the 11th FYP to the end of the 12th

FYP. Almost all the component years of these two FYPs were MPSS whatever the ratio of the shared inputs. That refers to the stability of the two sectors and their strong resistance to the impact of the local and global markets in the last two FYPs.

**Table 3 Industry's MPSS scores for k=9 partitions of shared inputs**

| FYPs | Years | Industry's MPSS scores, epsilon=0.1, nine partitions to the inputs | | | | | | | | |
| --- | --- | --- | --- | --- | --- | --- | --- | --- | --- | --- |
| | | k=1 | k=2 | k=3 | k=4 | k=5 | k=6 | k=7 | k=8 | k=9 |
| 3 | 1966 | 0.1871 | 0.0000 | 0.0000 | 0.0000 | 0.0000 | 0.0000 | 0.0000 | 0.0000 | 0.0000 |
| | 1967 | 15.480 | 15.750 | 15.480 | 15.480 | 15.480 | 15.480 | 15.480 | 15.480 | 15.480 |
| | 1968 | 28.418 | 31.781 | 28.418 | 28.418 | 31.781 | 28.418 | 20.138 | 31.781 | 20.138 |
| | 1969 | 17.177 | 18.417 | 17.177 | 18.417 | 17.177 | 17.177 | 17.177 | 17.177 | 5.5736 |
| | 1970 | 0.1093 | 0.1093 | 0.1093 | 0.1093 | 0.1093 | 0.1093 | 0.1093 | 0.1093 | 0.1093 |
| 4 | 1971 | 0.0362 | 0.1665 | 0.1665 | 2.6010 | 2.6010 | 2.6010 | 2.6010 | 2.6010 | 0.1665 |
| | 1972 | 0.1275 | 0.1275 | 0.1275 | 0.1275 | 0.1275 | 1.6013 | 1.6013 | 1.6013 | 1.6013 |
| | 1973 | 0.0452 | 0.0452 | 0.0452 | 0.0452 | 0.0452 | 0.0452 | 0.0452 | 0.0452 | 0.0452 |
| | 1974 | 0.1068 | 0.1078 | 0.1068 | 1.5042 | 1.5042 | 1.5042 | 1.5042 | 1.5042 | 1.5042 |
| | 1975 | 0.0000 | 0.0000 | 0.0000 | 0.0000 | 0.0000 | 0.0000 | 0.0000 | 0.0000 | 0.0000 |
| 5 | 1976 | 0.0928 | 0.0928 | 0.0803 | 0.9803 | 0.9803 | 0.9803 | 0.9803 | 0.9803 | 0.9803 |
| | 1977 | 0.0264 | 0.0166 | 0.0264 | 0.0264 | 0.0264 | 0.0264 | 0.0264 | 0.0264 | 0.0264 |
| | 1978 | 0.0255 | 0.0090 | 0.0255 | 0.0255 | 0.0255 | 0.0255 | 0.0255 | 0.0255 | 0.0255 |
| | 1979 | 1.9400 | 0.1085 | 1.9400 | 0.0165 | 0.0165 | 0.0165 | 0.0165 | 0.0165 | 0.0165 |
| | 1980 | 1.0212 | 0.0640 | 0.0979 | 0.0979 | 0.0979 | 0.0979 | 0.0979 | 0.0979 | 0.0979 |
| 6 | 1981 | 1.1090 | 0.0698 | 0.0969 | 0.0969 | 0.0969 | 0.0969 | 0.0969 | 0.0969 | 0.0969 |
| | 1982 | 2.0115 | 0.1099 | 2.0115 | 2.0115 | 0.0230 | 0.0230 | 0.0230 | 0.0230 | 0.0230 |
| | 1983 | 0.8806 | 1.4087 | 0.8806 | 0.0481 | 0.0481 | 0.0481 | 0.0481 | 0.0481 | 0.0481 |
| | 1984 | 5.3650 | 2.6671 | 1.9106 | 1.9106 | 1.9106 | 1.9106 | 1.9106 | 1.9106 | 1.9106 |
| | 1985 | 1.1554 | 0.5967 | 0.0268 | 0.0268 | 0.0268 | 0.0268 | 0.0268 | 0.0268 | 0.0268 |
| 7 | 1986 | 0.7613 | 0.7613 | 0.7613 | 0.0541 | 0.0541 | 0.0541 | 0.0541 | 0.0541 | 0.0541 |
| | 1987 | 0.0028 | 0.0028 | 0.0028 | 0.0028 | 0.0028 | 0.0028 | 0.0028 | 0.0028 | 0.0028 |
| | 1988 | 0.7215 | 0.7215 | 0.7215 | 0.7215 | 0.7215 | 0.7215 | 0.7215 | 0.7215 | 0.7215 |
| | 1989 | 0.4476 | 0.0252 | 0.0252 | 0.0252 | 0.0252 | 0.0252 | 0.0252 | 0.0252 | 0.0252 |
| | 1990 | 1.0160 | 1.0160 | 1.0160 | 1.0160 | 1.0160 | 1.0160 | 1.0160 | 1.0160 | 0.1081 |
| 8 | 1991 | 1.0505 | 1.0505 | 1.0505 | 1.0505 | 1.0505 | 1.0505 | 1.0505 | 1.0505 | 1.0505 |
| | 1992 | 0.8610 | 0.8610 | 0.8610 | 0.8610 | 0.8610 | 0.8610 | 0.8610 | 0.8610 | 0.8610 |
| | 1993 | 0.3360 | 0.3360 | 0.3360 | 0.3360 | 0.0676 | 0.0676 | 0.0676 | 0.0676 | 0.0676 |
| | 1994 | 0.1771 | 0.0299 | 0.0299 | 0.0299 | 0.0299 | 0.0299 | 0.0239 | 0.0000 | 0.0000 |
| | 1995 | 0.0143 | 0.0143 | 0.0143 | 0.0143 | 0.0143 | 0.0000 | 0.0000 | 0.0000 | 0.0000 |
| 9 | 1996 | 0.0597 | 0.0597 | 0.0597 | 0.0190 | 0.0000 | 0.0000 | 0.0000 | 0.0000 | 0.0000 |
| | 1997 | 0.2927 | 0.2927 | 0.2927 | 0.0000 | 0.0000 | 0.0000 | 0.0000 | 0.0000 | 0.0000 |
| | 1998 | 0.0096 | 0.0096 | 0.0000 | 0.0000 | 0.0000 | 0.0000 | 0.0000 | 0.0000 | 0.0000 |
| | 1999 | 0.0146 | 0.0146 | 0.0000 | 0.0000 | 0.0000 | 0.0000 | 0.0000 | 0.0000 | 0.0000 |
|  | 2000 | 0.0142 | 0.0181 | 0.0000 | 0.0000 | 0.0000 | 0.0000 | 0.0000 | 0.0000 | 0.0000 |
|  | 2001 | 0.0287 | 0.0000 | 0.0000 | 0.0000 | 0.0000 | 0.0000 | 0.0000 | 0.0000 | 0.0000 |
|  | 2002 | 0.0366 | 0.0000 | 0.0000 | 0.0000 | 0.0000 | 0.0000 | 0.0000 | 0.0000 | 0.0000 |
| 10 | 2003 | 0.0203 | 0.0000 | 0.0000 | 0.0000 | 0.0000 | 0.0000 | 0.0000 | 0.0000 | 0.0000 |
|  | 2004 | 0.0220 | 0.0000 | 0.0000 | 0.0000 | 0.0000 | 0.0000 | 0.0000 | 0.0000 | 0.0000 |
|  | 2005 | 0.0000 | 0.0000 | 0.0000 | 0.0000 | 0.0000 | 0.0000 | 0.0000 | 0.0000 | 0.0000 |
|  | 2006 | 0.0000 | 0.0000 | 0.0000 | 0.0000 | 0.0000 | 0.0000 | 0.0000 | 0.0000 | 0.0000 |
|  | 2007 | 0.0000 | 0.0000 | 0.0000 | 0.0000 | 0.0000 | 0.0000 | 0.0000 | 0.0000 | 0.0000 |
| 11 | 2008 | 0.0000 | 0.0000 | 0.0000 | 0.0000 | 0.0000 | 0.0000 | 0.0000 | 0.0000 | 0.0000 |
|  | 2009 | 0.0000 | 0.0000 | 0.0000 | 0.0000 | 0.0000 | 0.0000 | 0.0000 | 0.0000 | 0.0000 |
|  | 2010 | 0.0000 | 0.0000 | 0.0000 | 0.0000 | 0.0000 | 0.0000 | 0.0000 | 0.0000 | 0.0000 |
|  | 2011 | 0.0000 | 0.0000 | 0.0000 | 0.0000 | 0.0000 | 0.0000 | 0.0000 | 0.0000 | 0.0000 |
|  | 2012 | 0.0000 | 0.0000 | 0.0000 | 0.0000 | 0.0000 | 0.0000 | 0.0000 | 0.0000 | 0.0000 |
| 12 | 2013 | 0.0000 | 0.0000 | 0.0000 | 0.0000 | 0.0000 | 0.0000 | 0.0000 | 0.0000 | 0.0000 |
|  | 2014 | 0.0000 | 0.0000 | 0.0000 | 0.0000 | 0.0000 | 0.0000 | 0.0000 | 0.0000 | 0.0000 |
|  | 2015 | 0.0000 | 0.0000 | 0.0000 | 0.0000 | 0.0000 | 0.0000 | 0.0000 | 0.0000 | 0.0000 |
| **No.** |  | 12 | 17 | 20 | 21 | 22 | 23 | 23 | 24 | 24 |
| **Mean** |  | 1.6240 | 1.5372 | 1.4780 | 1.5215 | 1.5184 | 1.4803 | 1.3146 | 1.5470 | 1.0152 |
| **Min** |  | 0.0000 | 0.0000 | 0.0000 | 0.0000 | 0.0000 | 0.0000 | 0.0000 | 0.0000 | 0.0000 |
| **Max** |  | 28.418 | 31.781 | 28.418 | 28.418 | 31.780 | 28.418 | 20.138 | 31.780 | 20.138 |

Applying model (3), the Industry MPSS scores over the ten FYPs are reported in Table 4. In the 3$^{rd}$ FYP, only the first year is MPSS when the Industry sector shares at least 20% of the inputs. The next MPSS year is 1975, in the 4$^{th}$ FYP, for all partitions of inputs. Starting from the year 1995, the Industry sector became MPSS in each year for high ratios of shared inputs. Then these high ratios started to be small and small until the year 2005 where Industry became MPSS for all partitions of inputs and continued like this to the end of the year 2015. The last two FYPs, 11$^{th}$ and 12$^{th}$, were the perfect FYPs among the others in the Industry sector.

The Agriculture MPSS scores of the ten FYPs are depicted in Table 5. We can notice that the Agriculture MPSS years are a little different from those MPSS years of Industry and overall FYPs. The 3$^{rd}$ FYP has two years, 1967 and 1968, were MPSS for almost all partitions of inputs. In the 4$^{th}$ FYP, the year 1975 was the only MPSS year. There were some problems in the 5$^{th}$ FYP where no MPSS years are noticed. In the 6$^{th}$ FYP, three years were MPSS for all partitions of inputs followed with two decades with only two MPSS years. Starting from the year 2004, the Agriculture sector became MPSS in each year with a high ratio of shared inputs until the end of the year 2015. The 11$^{th}$ FYP was almost stable, and MPSS for almost partitions and the 12$^{th}$ was the perfect FYP among the then FYPs in the Agriculture sector.

**Table 4 Agriculture's MPSS scores for k=9 partitions of shared inputs**

| FYPs | Years | Agriculture's MPSS scores, epsilon=0.1, nine partitions to the inputs ||||||||| 
| | | k=1 | k=2 | k=3 | k=4 | k=5 | k=6 | k=7 | k=8 | k=9 |
|---|---|---|---|---|---|---|---|---|---|---|
| 3 | 1966 | 0.0888 | 0.0820 | 0.0820 | 0.0820 | 0.0820 | 0.0820 | 0.0820 | 0.0820 | 0.0980 |
| | 1967 | 0.0047 | 0.0000 | 0.0000 | 0.0000 | 0.0000 | 0.0000 | 0.0000 | 0.0000 | 0.0000 |
| | 1968 | 0.0000 | 0.0000 | 0.0000 | 0.0000 | 0.0000 | 0.0000 | 0.0000 | 0.0000 | 0.0000 |
| | 1969 | 0.0808 | 0.0808 | 0.0220 | 0.0220 | 0.0220 | 0.0220 | 0.0220 | 0.0220 | 0.0220 |
| | 1970 | 0.0580 | 0.0526 | 0.0526 | 0.0526 | 0.0526 | 0.0526 | 0.0526 | 0.0526 | 0.0573 |
| 4 | 1971 | 0.1058 | 0.0842 | 0.0842 | 0.0842 | 0.0842 | 0.0842 | 0.0842 | 0.0842 | 0.0842 |
| | 1972 | 0.0919 | 0.0866 | 0.0866 | 0.0866 | 0.0866 | 0.0866 | 0.0866 | 0.0866 | 0.0866 |
| | 1973 | 0.0033 | 0.0033 | 0.0007 | 0.0007 | 0.0007 | 0.0007 | 0.0007 | 0.0007 | 0.0007 |
| | 1974 | 0.0087 | 0.0087 | 0.0065 | 0.0065 | 0.0065 | 0.0065 | 0.0065 | 0.0065 | 0.0065 |
| | 1975 | 0.0000 | 0.0000 | 0.0000 | 0.0000 | 0.0000 | 0.0000 | 0.0000 | 0.0000 | 0.0000 |
| 5 | 1976 | 0.0152 | 0.0152 | 0.0130 | 0.0130 | 0.0130 | 0.0130 | 0.0130 | 0.0130 | 0.0130 |
| | 1977 | 0.1149 | 0.1149 | 0.0679 | 0.0679 | 0.0679 | 0.0679 | 0.0679 | 0.0679 | 0.0679 |
| | 1978 | 0.0846 | 0.1999 | 0.1794 | 0.1794 | 0.1794 | 0.1794 | 0.1794 | 0.1794 | 0.1794 |
| | 1979 | 0.0689 | 0.0689 | 0.0689 | 0.0689 | 0.0659 | 0.0659 | 0.0659 | 0.0659 | 0.0659 |
| | 1980 | 0.0819 | 0.0819 | 0.0819 | 0.0819 | 0.0819 | 0.0783 | 0.0783 | 0.0783 | 0.0783 |
| 6 | 1981 | 0.0562 | 0.0562 | 0.0410 | 0.0559 | 0.0559 | 0.0559 | 0.0559 | 0.0559 | 0.0559 |
| | 1982 | 0.0000 | 0.0000 | 0.0000 | 0.0000 | 0.0000 | 0.0000 | 0.0000 | 0.0000 | 0.0000 |
| | 1983 | 0.0000 | 0.0000 | 0.0000 | 0.0000 | 0.0000 | 0.0000 | 0.0000 | 0.0000 | 0.0000 |
| | 1984 | 0.0000 | 0.0000 | 0.0000 | 0.0000 | 0.0000 | 0.0000 | 0.0000 | 0.0000 | 0.0000 |
| | 1985 | 0.0205 | 0.0075 | 0.0075 | 0.0075 | 0.0075 | 0.0075 | 0.0075 | 0.0075 | 0.0075 |
| 7 | 1986 | 0.0684 | 0.0667 | 0.0667 | 0.0667 | 0.0667 | 0.0667 | 0.0667 | 0.0667 | 0.0667 |
| | 1987 | 0.0772 | 0.0772 | 0.0772 | 0.0772 | 0.0772 | 0.0450 | 0.0450 | 0.0450 | 0.0772 |
| | 1988 | 0.0506 | 0.0506 | 0.0506 | 0.0506 | 0.0506 | 0.0350 | 0.0350 | 0.0350 | 0.0506 |
| | 1989 | 0.0709 | 0.0709 | 0.0669 | 0.0669 | 0.0669 | 0.0669 | 0.0669 | 0.0669 | 0.0669 |
| | 1990 | 0.0037 | 0.0037 | 0.0037 | 0.0037 | 0.0037 | 0.0037 | 0.0037 | 0.0037 | 0.0037 |
| 8 | 1991 | 0.0819 | 0.0819 | 0.0727 | 0.0727 | 0.0727 | 0.0727 | 0.0727 | 0.0727 | 0.0727 |
| | 1992 | 0.0181 | 0.0181 | 0.0119 | 0.0119 | 0.0119 | 0.0119 | 0.1425 | 0.1425 | 0.1425 |
| | 1993 | 0.0846 | 0.0846 | 0.0846 | 0.0846 | 0.0846 | 0.0846 | 0.2293 | 0.2293 | 0.2293 |
| | 1994 | 0.0485 | 0.0485 | 0.0422 | 0.0422 | 0.0422 | 0.0422 | 0.0422 | 0.1093 | 0.1093 |
| | 1995 | 0.0100 | 0.0100 | 0.0000 | 0.0000 | 0.0000 | 0.0000 | 0.0000 | 0.0000 | 0.0000 |
| 9 | 1996 | 0.0000 | 0.0000 | 0.0000 | 0.0000 | 0.0000 | 0.0000 | 0.0000 | 0.0000 | 0.0000 |
| | 1997 | 0.0216 | 0.0216 | 0.0216 | 0.0216 | 0.0216 | 0.0216 | 0.0216 | 0.0216 | 0.0216 |
| | 1998 | 0.0295 | 0.0295 | 0.0295 | 0.0295 | 0.0295 | 0.0295 | 0.0295 | 0.0295 | 0.0295 |
| | 1999 | 0.0658 | 0.0658 | 0.0658 | 0.0658 | 0.0658 | 0.0658 | 0.0047 | 0.0047 | 0.0047 |
| | 2000 | 0.1094 | 0.1094 | 0.1094 | 0.1094 | 0.1094 | 0.1094 | 0.0492 | 0.0492 | 0.0492 |
| 10 | 2001 | 0.1111 | 0.1111 | 0.1111 | 0.1111 | 0.1111 | 0.1111 | 0.1111 | 0.0547 | 0.0547 |
| | 2002 | 0.1199 | 0.1199 | 0.1199 | 0.1199 | 0.1199 | 0.1199 | 0.1199 | 0.0664 | 0.0664 |

| | Year | | | | | | | | | |
|---|---|---|---|---|---|---|---|---|---|---|
| | 2003 | 0.1481 | 0.1481 | 0.1481 | 0.1481 | 0.1481 | 0.1481 | 0.1481 | 0.0947 | 0.0947 |
| | 2004 | 0.0000 | 0.0000 | 0.0356 | 0.0356 | 0.0356 | 0.0356 | 0.0356 | 0.0356 | 0.0322 |
| | 2005 | 0.0000 | 0.0000 | 0.0852 | 0.0852 | 0.0852 | 0.0852 | 0.0852 | 0.0852 | 0.0808 |
| | 2006 | 0.0000 | 0.0000 | 0.1141 | 0.1141 | 0.1141 | 0.1141 | 0.1141 | 0.1141 | 0.1141 |
| | 2007 | 0.0000 | 0.0000 | 0.0000 | 0.0000 | 0.0000 | 0.0752 | 0.0752 | 0.0752 | 0.0752 |
| 11 | 2008 | 0.0000 | 0.0000 | 0.0000 | 0.0000 | 0.0000 | 0.0000 | 0.0000 | 0.0009 | 0.0009 |
| | 2009 | 0.0000 | 0.0000 | 0.0000 | 0.0000 | 0.0000 | 0.0000 | 0.0000 | 0.0000 | 0.0253 |
| | 2010 | 0.0000 | 0.0000 | 0.0000 | 0.0000 | 0.0000 | 0.0000 | 0.0000 | 0.0000 | 0.0021 |
| | 2011 | 0.0000 | 0.0000 | 0.0000 | 0.0000 | 0.0000 | 0.0000 | 0.0000 | 0.0000 | 0.0000 |
| | 2012 | 0.0000 | 0.0000 | 0.0000 | 0.0000 | 0.0000 | 0.0000 | 0.0000 | 0.0000 | 0.0000 |
| 12 | 2013 | 0.0000 | 0.0000 | 0.0000 | 0.0000 | 0.0000 | 0.0000 | 0.0000 | 0.0000 | 0.0000 |
| | 2014 | 0.0000 | 0.0000 | 0.0000 | 0.0000 | 0.0000 | 0.0000 | 0.0000 | 0.0000 | 0.0000 |
| | 2015 | 0.0000 | 0.0000 | 0.0000 | 0.0000 | 0.0000 | 0.0000 | 0.0000 | 0.0000 | 0.0000 |
| **No.** | | 18 | 19 | 17 | 17 | 17 | 16 | 16 | 15 | 13 |
| **Mean** | | 0.0401 | 0.0412 | 0.0422 | 0.0425 | 0.0425 | 0.0429 | 0.0460 | 0.0441 | 0.0459 |
| **Min** | | 0.0000 | 0.0000 | 0.0000 | 0.0000 | 0.0000 | 0.0000 | 0.0000 | 0.0000 | 0.0000 |
| **Max** | | 0.1481 | 0.1999 | 0.1794 | 0.1794 | 0.1794 | 0.1794 | 0.2293 | 0.2293 | 0.2293 |

From Tables (5), it is evident that the decomposition theorem of parallel network MPSS, Theorem 1, is satisfied in each year.

Figure 3 displays the growth in the productivity scale size of the aggregated MPSS years of overall FYP, Industry, and Agriculture for the nine partitions of the resources. Industry needs 25% of the shared resources to have the best economic scale size compared to Agriculture. For an equal ratio of partitions (k=5; 50 Industry: 50 Agriculture), the Industry sector is MPSS in 22 years compared to 17 years in Agriculture. In other words, using the same resources of population, GDP, and general government final consumption, the Industry sector has better economic scale than the Agriculture sector. Only when the Agriculture sector shares at least 80% of the inputs, it has better economic scale than the Industry. In average, the overall FYP is MPSS in 11 years.

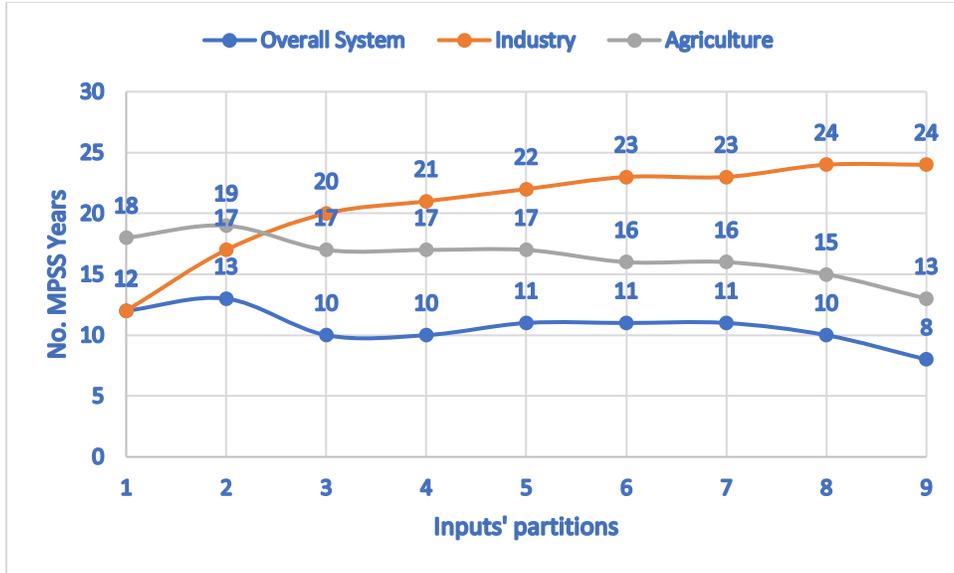

**Figure 2 MPSS years of the overall system, Industry, and Agriculture**

For k=5, the two sectors share the same amount of the resources. Figure 4 displays the number of MPSS years in each FYP for the Agriculture and Industry sectors and the overall system as well.

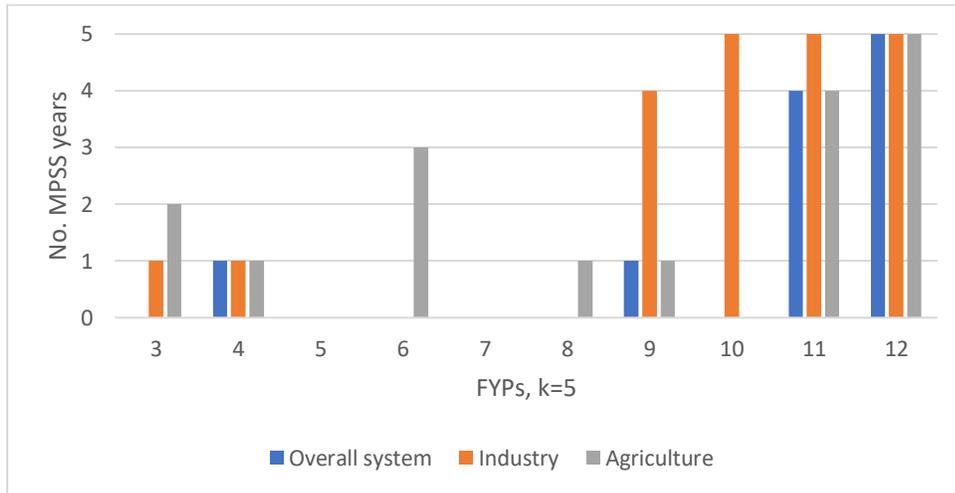

**Figure 3 MPSS years of China's FYPs for k=5**

Industry sector started to have the best economic scale size from the 9th FYP. The 10th and 11th FYPs were MPSS in the overall and the Agriculture and Industry sectors.

## 4 Conclusions and implications

In this study, we defined the MPSS concept in classical parallel networks and then in parallel networks with shared inputs. A new relational model for measuring the MPSS of the overall system and the internal subsystems is proposed. Mathematical analysis proved that the system MPSS could be decomposed as the weighted sum of the MPSS values of the internal subsystems. As a result, the overall system is said to be MPSS if and only if it is MPSS in each internal subsystem.

A real-life application of China's Five-Year Plans with shared inputs is used to show the applicability and the merits of the proposed models and theorems. Exciting findings have been noticed. First, using at least 25% of the shared inputs enable the Industry sector to achieve MPSS more than Agriculture. Second, for an equal ratio of partitions (k=5; 50 Industry: 50 Agriculture), the Industry sector was MPSS in 22 years compared to 17 years in Agriculture. In other words, using the same resources of population, GDP, and general government final consumption, the Industry sector has better economic scale than the Agriculture sector. Third, only when the Agriculture sector shares at least 80% of the inputs, it has better economic scale than the Industry. Furthermore, the last two FYPs, 11$^{th}$ and 12$^{th}$, were the perfect two FYPs among the others.

The importance of this work comes from the fact that the most productive scale size of decision making units is an important topic and not studied in network DEA before. This paper could introduce the MPSS concept to network DEA with parallel structure. The policymakers can get many clear and accurate insights into the productivity scale size of their organizations. Furthermore, the place of inefficient scale size within the whole system will be readily determined and improved. Improvement strategies for the evaluated decision making units are proposed and suggested.

## 4.1 Limitations

A limitation of this paper is that the proposed network MPSS models are in the dual form, and thus assuming restrictions or giving weights for some inputs or outputs is not possible. In other words, assurance region (AR) cannot be implemented in the MPSS models. As we noticed in this study, the relationship between the overall MPSS of the system is the weighted sum of the internal processes MPSSs. The importance of each process cannot be calculated by the MPSS model but selected by the decision makers.

## 4.2 Future research

This work could investigate the concept of MPSS in networks with parallel structure. However, there are still different network structures should be considered for MPSS evaluation such as networks with the mixed structure of series and parallel. In DEA literature, standard most productive scale size (MPSS) model maximizes the average of the difference of inputs and outputs of a production system in a specified period, where variations in different periods are ignored. It would be interesting if we could take the operations of individual periods into account and develop a multi-period MPSS model, dynamic MPSS model, to measure the overall MPSS and period MPSSs at the same time.

Yang, F., Du, F., Liang, L., & Yang, Z. (2014). Forecasting the Production Abilities of Recycling Systems: A DEA Based Research. *Journal of Applied Mathematics*, *2014*, 1–9. https://doi.org/10.1155/2014/961468

Yang, M., & Yang, F. (2016). Energy-Efficiency Policies and Energy Productivity Improvements: Evidence from China's Manufacturing Industry. *Emerging Markets Finance and Trade*, *52*(6), 1395–1404. https://doi.org/10.1080/1540496X.2016.1152800

Zhu, J. (2000). Further discussion on linear production functions and DEA. *European Journal of Operational Research*, *127*(3), 611–618. https://doi.org/10.1016/S0377-2217(99)00344-6